\documentclass[aps,prl,twocolumn,nopacs,superscriptaddress]{revtex4}

\usepackage{graphicx}
\usepackage{color}
\usepackage{amsfonts}
\begin{document}

\title{Transfer-free electrical insulation of epitaxial graphene from its metal substrate}

\author{Silvano~Lizzit}
\affiliation{Sincrotrone Trieste, S.S. 14 Km 163.5, 34149
Trieste, Italy}
\author{Rosanna Larciprete}
\affiliation{CNR-Institute for Complex Systems, Via Fosso del Cavaliere 100, 00133 Roma, Italy}
\author{Paolo Lacovig}
\affiliation{Sincrotrone Trieste, S.S. 14 Km 163.5, 34149
Trieste, Italy}
\author{Matteo Dalmiglio}
\affiliation{Sincrotrone Trieste, S.S. 14 Km 163.5, 34149
Trieste, Italy}
\author{Fabrizio Orlando}
\affiliation{Physics Department and CENMAT, University of Trieste, Via Valerio 2, 34127 Trieste, Italy}
\affiliation{IOM-CNR Laboratorio TASC, Area Science Park, 34149 Trieste, Italy}
\author{Alessandro Baraldi}
\affiliation{Physics Department and CENMAT, University of Trieste, Via Valerio 2, 34127 Trieste, Italy}
\affiliation{IOM-CNR Laboratorio TASC, Area Science Park, 34149 Trieste, Italy}
\author{Lauge Gammelgaard}
\affiliation{Capres A/S, 2800 Kgs. Lyngby, Denmark}
\author{Lucas Barreto}
\affiliation{Department of Physics and Astronomy, Interdisciplinary Nanoscience Center, Aarhus University,
8000 Aarhus C, Denmark}
\author{Marco Bianchi}
\affiliation{Department of Physics and Astronomy, Interdisciplinary Nanoscience Center, Aarhus University,
8000 Aarhus C, Denmark}
\author{Edward Perkins}
\affiliation{Department of Physics and Astronomy, Interdisciplinary Nanoscience Center, Aarhus University,
8000 Aarhus C, Denmark}
\author{Philip~Hofmann}
\affiliation{Department of Physics and Astronomy, Interdisciplinary Nanoscience Center, Aarhus University,
8000 Aarhus C, Denmark}
\email{philip@phys.au.dk}

%%%%%%%%%%%%%%%%%%%%%%%%%%%%%%%%%%%%%%%%%%%%%%%%%%%%%%%%%%%%%%%%%%%%%
%% The document title should be given as usual
%% A short title can be given as a *suggestion* for running headers.
%%%%%%%%%%%%%%%%%%%%%%%%%%%%%%%%%%%%%%%%%%%%%%%%%%%%%%%%%%%%%%%%%%%%%

%%%%%%%%%%%%%%%%%%%%%%%%%%%%%%%%%%%%%%%%%%%%%%%%%%%%%%%%%%%%%%%%%%%%%
%% The manuscript does not need to include \maketitle, which is
%% executed automatically.  The document should begin with an
%% abstract, if appropriate.  If one is given and should not be, the
%% contents will be gobbled.
%%%%%%%%%%%%%%%%%%%%%%%%%%%%%%%%%%%%%%%%%%%%%%%%%%%%%%%%%%%%%%%%%%%%%
\begin{abstract}
High-quality, large-area epitaxial graphene can be grown on metal surfaces but its transport properties cannot be exploited because the electrical conduction is dominated by the substrate. Here we insulate epitaxial graphene on Ru(0001) by a step-wise intercalation of silicon and oxygen, and the eventual formation of a SiO$_2$ layer between the graphene and the metal. We follow the reaction steps by x-ray photoemission spectroscopy and demonstrate the electrical insulation using a nano-scale multipoint probe technique. 
\end{abstract}

\maketitle
Graphene, a single layer of carbon atoms \cite{Novoselov:2005,Geim:2007}, is one of the most promising materials for future electronic applications because of its very high carrier mobility, tolerance to high temperatures and inertness \cite{Kim:2011b,Avouris:2010}. Exploiting the electronic properties requires graphene to be placed on an insulating substrate, such as SiO$_2$. This can be achieved by different routes, for example after exfoliation from graphite \cite{Novoselov:2004}, after reduction of graphene oxide \cite{Park:2009c,Gomez:2010} or after large-scale growth on metal films that are subsequently dissolved  \cite{Bae:2010}, evaporated  \cite{Ismach:2010}or removed by peeling \cite{Su:2011}. Unfortunately, these methods result either in very small graphene flakes or in graphene of poor quality. Moreover, the transfer process itself introduces large amounts of defects in the graphene lattice, inevitably leading to a strong decrease in carrier mobilities. A proven route to large-scale, single-layer graphene is the epitaxial growth on transition metal surfaces  \cite{Sutter:2008,Vazquez-de-Parga:2008,Gunther:2011} but this has the disadvantage of a conductive substrate, rendering the conduction through graphene irrelevant. Here we demonstrate a transfer-free method to electrically insulate such epitaxial graphene from the metal it is grown on. This is achieved by growing an insulating layer of SiO$_2$ of the desired thickness directly under the graphene layer, through a stepwise reaction between intercalated silicon and oxygen. We show that in this system the transport is dominated by graphene and not by the underlying metal. This route combines the advantages of high-quality large area graphene growth with an insulating substrate, opening new perspectives for device fabrication and fundamental studies of transport properties.

\begin{figure}
\begin{center}
\includegraphics[width=\columnwidth]{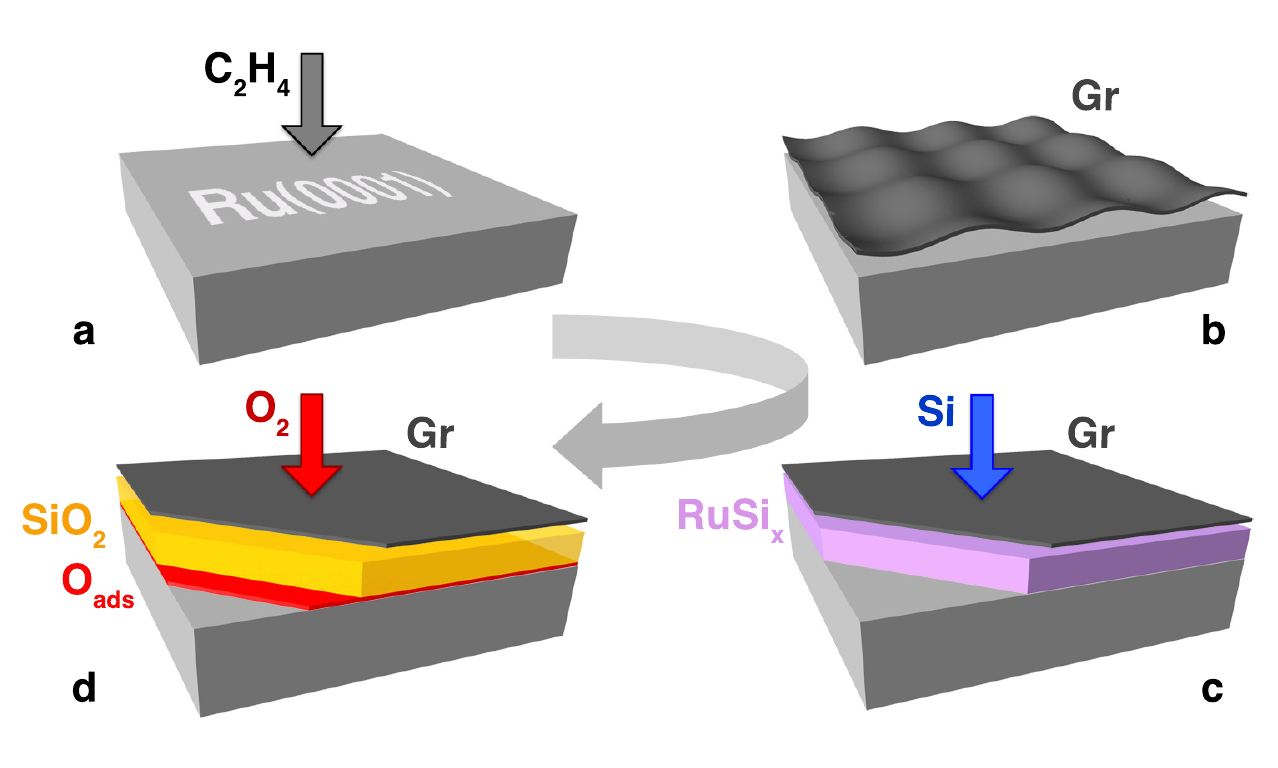}
\caption{Synthesis of SiO$_2$ under epitaxial graphene on Ru(0001). Each process step is  schematically illustrated. (a) C$_2$H$_4$ is adsorbed on a Ru single crystal surface and decomposed at high temperature, leading to (b) the formation of epitaxial graphene. (c) The sample surface is exposed to silicon that intercalates below the graphene and forms a silicide with the metal substrate. (d) The metal silicide is oxidized to form an insulating SiO$_2$ layer that separates the metal from graphene. The metal surface is terminated with chemisorbed oxygen.
\label{fig:1}}
\end{center}
\end{figure}

The procedure used to insulate epitaxial graphene from its metal substrate is schematically outlined in  Fig. \ref{fig:1}. Epitaxial graphene is grown on a clean Ru(0001) crystal surface. The graphene layer is then exposed to silicon that intercalates below the graphene \cite{Wang:2011d,Xia:2012,Mao:2012} and forms a silicide with the metal substrate. A subsequent exposure to oxygen also leads to an intercalation and an oxidation of the metal silicide, resulting in an insulating SiO$_2$ layer that separates the metal from graphene. The thickness of the SiO$_2$ can be varied by the amount of intercalated Si. We can follow each step of the process taking place under the graphene by high-resolution x-ray photoemission spectroscopy (XPS) and demonstrate the electrical insulation using a nanoscale four point probe technique.

\begin{figure*}
\begin{center}
\includegraphics[scale=0.7]{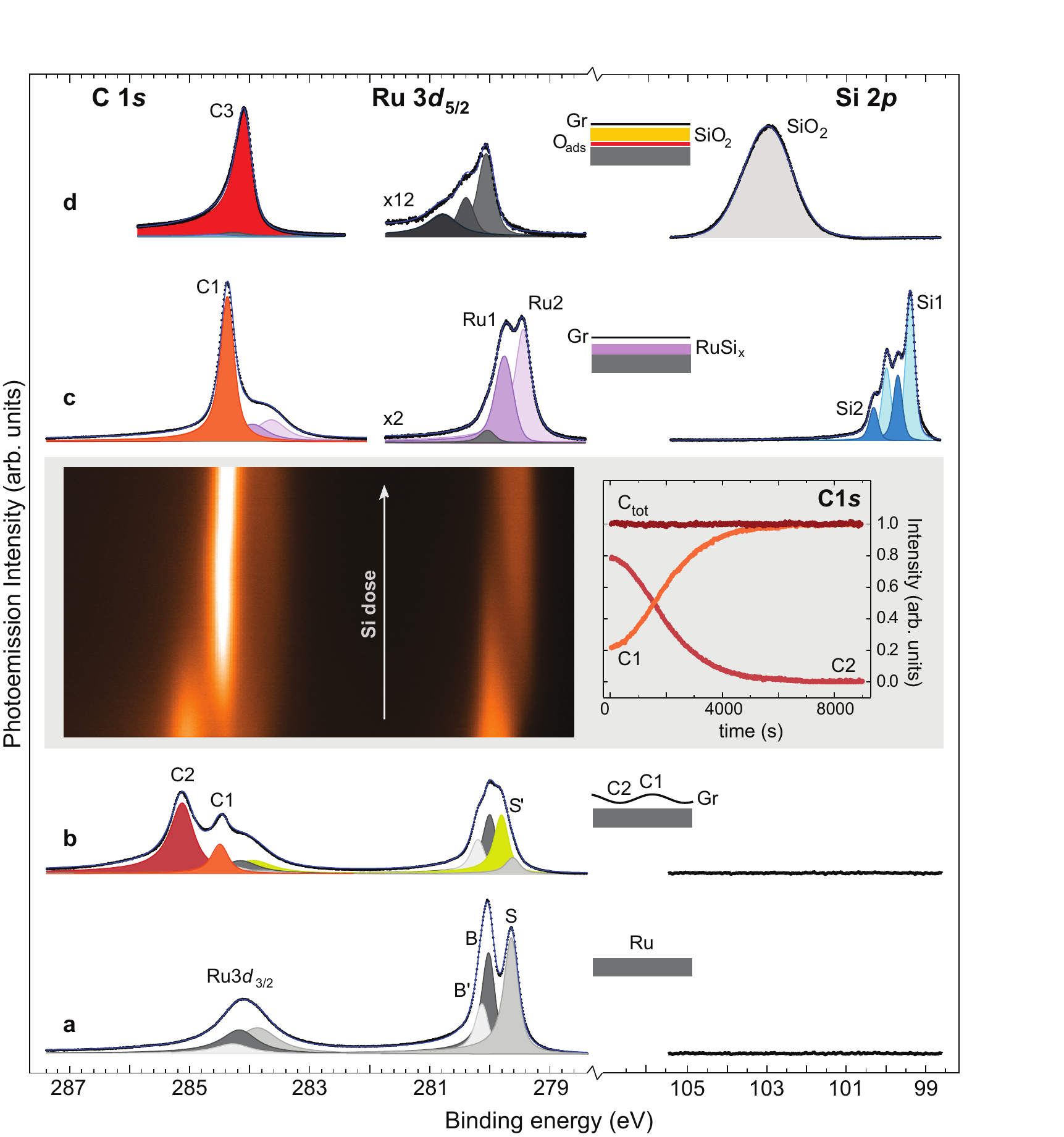}%, height=10cm]{fig2.pdf} %[width=\columnwidth]
\caption{Following the graphene formation and SiO$_2$ synthesis on Ru(0001) through XPS of the C~1s, Ru~3d and Si~2p core levels. (a) Core level spectra measured on the clean Ru(0001) surface.  The Ru~3d$_{5/2}$ peak shows the bulk component $B$ (280.01~eV) and the components due to Ru atoms in the first ($S$, 279.63~eV) and in the second ($B^{\prime}$, 280.12~eV) layer. (b) One layer of graphene on Ru(0001) shows two C~1s peaks for carbon strongly (C2, 285.11~eV) and weakly (C1, 284.48~eV) interacting with the Ru surface. The Ru~3d$_{5/2}$ shows the additional $S^{\prime}$ component (279.92~eV). (c) Intercalation of 4 ML of Si that alloys with the metal forming Ru silicide, giving rise to two Ru~3d$_{5/2}$ peaks (Ru1, 279.72~eV and Ru2, 279.41~eV), a single C~1s peak C1 and two Si~2p doublets with the $3/2$ peaks at 99.40 and 99.71~eV. (d) The oxidation of the Ru silicide leads to a single C~1s component (C3, 284.04~eV) shifted in binding energy from C1, a Ru~3d$_{5/2}$ spectrum characteristic for the oxygen-covered surface \cite{Lizzit:2001} and a broad Si~2p peak characteristic for SiO$_2$. The central inset displays a 2D plot of the fast XPS spectra measured while evaporating Si on graphene at 720~K, showing the decay of C1 and the rise of C2 component. The right part of the inset shows the evolution of both components as well as the sum of the intensity that remains constant in the intercalation process.
\label{fig:2}}
\end{center}
\end{figure*}

The starting point is a clean Ru(0001) single crystal surface. The Ru 3d XPS spectrum in  Fig. \ref{fig:2}a shows two spin-orbit split components. The higher binding energy component (Ru~3d$_{3/2}$) has a substantially shorter lifetime and is thus rather broad  \cite{Lizzit:2001}, but the lower binding energy component  (Ru~3d$_{5/2}$) is sufficiently sharp to show a shifted component $S$ on the low binding energy side of the bulk component $B$, caused by the atoms in the first layer of the crystal. A component $B^{\prime}$ due to Ru atoms in the second layer can be also distinguished. This spectrum is characteristic for the clean surface \cite{Lizzit:2001}.
In the next step, graphene is grown epitaxially, employing the standard approach of decomposing small hydrocarbon molecules at high temperature \cite{Vazquez-de-Parga:2008,Gunther:2011}. The C~1s and the  Ru~3d$_{3/2}$ peaks overlap energetically in the XPS spectrum. Nevertheless, a deconvolution provides detailed information on the corrugation of the graphene layer  \cite{Preobrajenski:2008b} ( Fig. \ref{fig:2}b). The  C~1s spectrum is dominated by the higher binding energy component (C2) that signals strong interaction with  the Ru atoms. The smaller, low binding energy peak (C1) corresponds to graphene areas weakly bound to the metal. The presence of the graphene layer is also reflected in an additional surface-atom peak  $S^{\prime}$.

In the next step, the surface is exposed to silicon at 720 K \cite{Mao:2012}. A plot of the C~1s and  Ru~3d spectral intensities versus Si dose is shown in the inset of  Fig. \ref{fig:2}. The intensity of the strongly interacting C~1s component C2 is gradually and completely transferred to the weakly interacting and narrow component C1. During this process, the  total C~1s intensity remains constant. This shows that all the Si atoms are intercalated below graphene and do not clusterize on top of it. In parallel, the Ru~3d$_{5/2}$ core level looses the $S$ and $S^{\prime}$ components characteristic of the Ru-graphene interface and shows two new components Ru1  and Ru2, shifted to lower binding energies, clearly indicating the formation of Ru silicide. \cite{Lu:1991, Pasquali:2009}. The intensity of the $B$ peak is very low, suggesting that the silicide formation extends over several layers. The spin-orbit split Si~2p core level measured after Si intercalation and  shown in  Fig. \ref{fig:2}c exhibits two doublet components Si1 and Si2, due to the formation of Ru-Si bonds and consistent with the Ru~3d$_{5/2}$ components Ru1 and Ru2. The total quantity of intercalated silicon atoms can be varied. In the experiment presented here, it corresponds to approximately 4 ML (1 ML=1 monolayer= 1.4 $\times$ 10$^{15}$ atoms/cm$^{2}$).

In a final step, the surface is exposed to molecular oxygen at pressure of $\approx4\times$10$^{-3}$~ mbar and temperature of 640 K. This results in the intercalation of oxygen below graphene \cite{Zhang:2009g,Sutter:2010} and in the progressive oxidation of the silicide layer. During the oxidation, the silicide components of the Si~2p spectrum are consumed and transformed into a broad peak at higher binding energy, indicative of Si in a SiO$_2$ environment \cite{Himpsel:1988} (Fig. \ref{fig:2}d). We find that SiO$_2$ is formed with a constant rate that is higher than that measured for the dry oxidation of the Si(001) surface with comparable temperature and O$_2$ pressure \cite{Enta:2008}. The rate-enhanced SiO$_2$ growth is an intrinsic advantage offered by the oxidation of the intermediate Ru silicide that forms at the metal surface during Si intercalation. The C~1s intensity converts into the single and narrow C3 peak, interpreted as graphene supported on SiO$_2$. Neither the Ru~3d$_{5/2}$ nor the C~1s spectra show any sign of oxidation, demonstrating that graphene does not react with O$_2$ during intercalation and that during the decomposition of the Ru silicide oxygen binds exclusively to silicon \cite{Abbati:1982, Cros:1983}. When the silicide is fully decomposed, the Ru~3d$_{5/2}$ spectrum shows the components characteristic for oxygen-covered Ru(0001)
 \cite{Lizzit:2001}. At this point the graphene is separated from the Ru crystal surface by  $\approx 1.8$~nm of SiO$_2$.

\begin{figure}
\begin{center}
\includegraphics[width=\columnwidth]{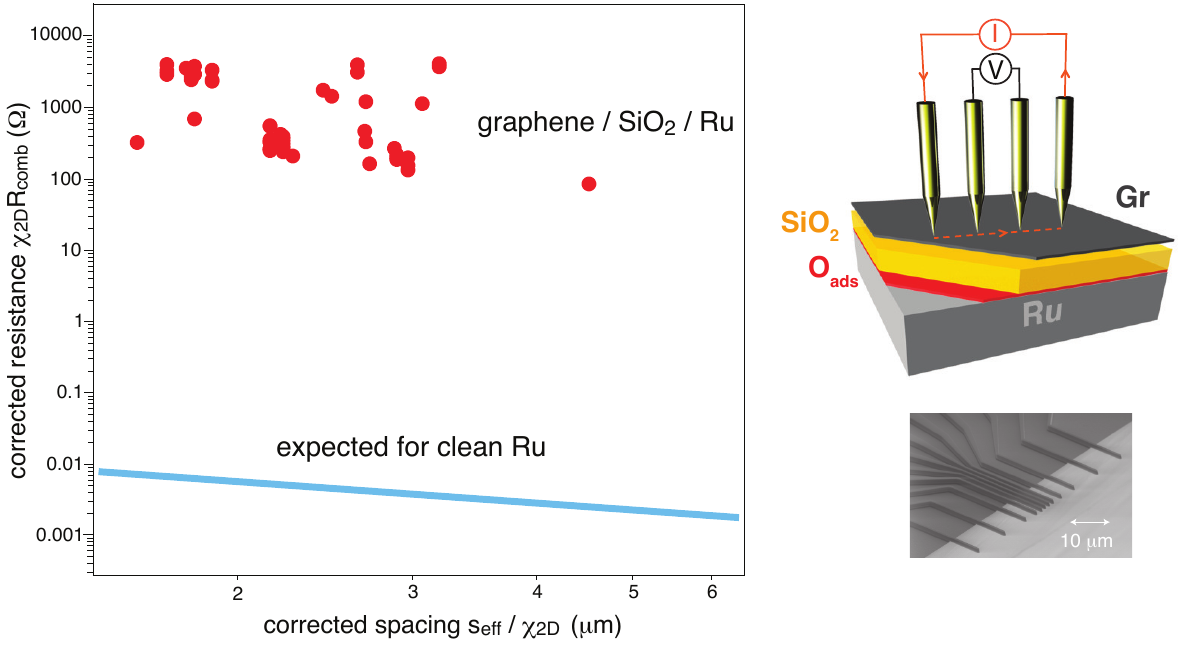}
\caption{Surface resistance measurements of the resulting graphene/SiO$_2$/Ru(0001) system using a micro four point probe. Shown is the corrected resistance $\chi_{2D} R_{comb}$ as a function of relative sensitivity ($s_{eff}/\chi_{2D}$). This essentially corresponds to the resistance as a function of contact spacing for a conventional four point probe measurement with four equidistant contacts but the definition of  $\chi_{2D} R_{comb}$ and $s_{eff}/\chi_{2D}$ permits measurements with non-equidistant contact spacings to be included in the plot. The blue line is the result that would be expected from a clean Ru surface or for a graphene layer on clean Ru. The inset shows a simple sketch of the measurement setup as well as a picture of the type of probe used for this experiment \cite{capres}.
 \label{fig:3}}
\end{center}
\end{figure}

This SiO$_2$ layer should now provide the electrical insulation of graphene and we proceed to experimentally test this. To this end, we perform a lateral transport measurement on the surface, using a microscopic 12 point probe \cite{capres}, shown in the inset of  Fig. \ref{fig:3}. Surface-sensitive transport can be achieved with four closely spaced contacts because the spreading of the current is confined to a depth comparable to the distance of the contacts \cite{Hofmann:2009}. More precisely, the expected measured four point probe resistances for a two-dimensional and a semi-infinite three-dimensional sample are $R_{2D}=\ln 2 / \pi \sigma_s$ and $R_{3D}=1/2\pi s \sigma_b$, respectively, where $\sigma_s$ is the sheet conductance, $\sigma_b$ the bulk conductance and $s$ the contact spacing. If we view the two-dimensional graphene and the three-dimensional substrate as two parallel resistors, graphene-dominated transport can be achieved for a sufficiently small contact spacing $s$ because then $R_{2D} \ll R_{3D}$. For graphene placed directly on a clean metal surface this is not possible (unless the mechanical contact is made only to the graphene and not to the metal) because a simple estimate of the required contact spacing results in an unachievably small value  (in the order of an atomic spacing) \cite{Sutter:2008}. Indeed, the same consideration implies that the construction of graphene electronics on a metal substrate is not a viable option.

With the highly resistive SiO$_2$ in between graphene and the metal, however, this situation changes.   Fig. \ref{fig:3}  shows the measured four point probe resistance on the SiO$_2$-intercalated graphene on Ru surface as a function of contact spacing. More precisely, the figure shows the corrected resistance $\chi_{2D} R_{comb}$ as a function of relative sensitivity ($s_{eff}/\chi_{2D}$), a transformation of the data that allows us to plot data measured with un-equal contact spacings as if it was measured with equal contact spacing. The measured resistance is roughly independent of the contact spacing, suggesting two-dimensional transport, and the resistance has the right order of magnitude for epitaxial graphene \cite{Jobst:2010} or exfoliated graphene placed on SiO$_2$, both measured with a lithographically fabricated device \cite{Zhang:2005} or with a four point probe similar to ours \cite{Klarskov:2011}. We attribute the considerable spread of the data points to a small non-uniformity of the SiO$_2$ film.  Most importantly, the measured resistance is about five orders of magnitude above the value one would expect to measure on a clean Ru(0001) surface in this region of contact spacings (indicated as a blue line), conclusively showing that the observed transport is not dominated by the substrate but by graphene.

The demonstrated process to insulate graphene from the metallic substrate it is grown on relies on the tendency of almost any adsorbate to intercalate under the graphene layer \cite{Wang:2011d,Xia:2012,Mao:2012,Zhang:2009g, Sutter:2010,Riedl:2009,Enderlein:2010,Emtsev:2011}, and this can be exploited to promote the chemical synthesis of materials below graphene. This proven concept opens  many design options and might thus have wide application in graphene research and device fabrication.

We thank the Carlsberg foundation for financial support. AB acknowledges the Universit\'{a} degli Studi di Trieste for the Finanziamento per Ricercatori di Ateneo. RL thanks the support of the COST Action MP0901 "NanoTP".
%
%\bibliographystyle{apsrev}
%\bibliography{groupreferences_new,locref}
%

\end{document}